\documentclass[amsmath,amssymb,showpacs,floatfix,twocolumn,aps,pra,10pt,superscriptaddress]{revtex4}

\usepackage{graphicx}

\def\iom{i\omega}
\def\ta{\mathrm{t}_\alpha}
\def\tl{\mathrm{t}_\ell}
\def\th{\mathrm{t}_\textit{h}}
\def\Va{\mathrm{V}_\alpha}
\def\Vl{\mathrm{V}_\ell}
\def\Vh{\mathrm{V}_\textit{h}}
\def\Za{Z_\alpha}
\def\Zl{Z_\ell}
\def\Zh{Z_\textit{h}}
\def\Da{\mathrm{D}_{\alpha}}
\def\Dl{\mathrm{D}_\ell}
\def\Dh{\mathrm{D}_\textit{h}}

\begin{document}

\title{Mott transition of fermionic mixtures with mass imbalance in optical lattices}
\date{\today}
\author{Tung-Lam Dao}
\affiliation{Laboratoire Charles Fabry de l'Institut d'Optique,
CNRS, Universit\'e Paris-Sud, Campus de l'{\'E}cole Polytechnique, 91127 Palaiseau Cedex, France}
\author{Michel Ferrero}
\affiliation{Centre de Physique Th\'eorique, \'Ecole
Polytechnique, CNRS, 91128 Palaiseau Cedex, France.}
\author{Pablo S. Cornaglia}
\affiliation{Centro At{\'o}mico Bariloche and Instituto Balseiro, CNEA, 8400 Bariloche, Argentina and\\
Consejo Nacional de Investigaciones Cient\'{\i}ficas y T\'ecnicas (CONICET), Argentina}
\author{Massimo Capone}
\affiliation{Consiglio Nazionale delle Ricerche, Istituto Officina dei Materiali (CNR/IOM), Uos Democritos, SISSA, International School for Advanced Studies (SISSA/ISAS), Via Bonomea 265, 34136 Trieste, Italy, and Dipartimento di Fisica, Universit\`a Sapienza, Piazzale Aldo Moro 2, Roma}

\begin{abstract}
We investigate the effect of mass imbalance in binary Fermi mixtures loaded in
optical lattices. Using dynamical mean-field theory, we study the transition
from a fluid to a Mott insulator driven by the repulsive interactions. For almost
every value of the parameters we find that the light species with smaller bare
mass is more affected by correlations than the heavy one, so that their
effective masses become closer than their bare masses before a Mott transition
occurs. The strength of the critical repulsion decreases monotonically as the
mass imbalance grows so that the minimum is realized when one of the species is
localized. The evolution of the spectral functions testifies that a continuous loss
of coherence and a destruction of the Fermi liquid occur as the imbalance grows. The two
species display distinct properties and experimentally-observable deviations
from the behavior of a balanced Fermi mixture.
\end{abstract}


\maketitle

\section{Introduction}
\label{sec:intro}

The ability to trap ultracold atomic gases and load them in optical lattices has led to the tantalizing possibility to create in a laboratory almost ideal realizations of popular condensed-matter models and to observe their remarkable properties~\cite{Jaksch:2005AnPhy.315}. One after the other, the cold atoms' route has met many fascinating milestones like the observation of the
superfluid to Mott-insulator transition in a system of bosons on a lattice~\cite{M.Greiner:Nature415}, the direct
imaging of the Fermi surface in a degenerate Fermi
gas~\cite{Kohl:PhysRevLett.94.080403}, the demonstration of
superfluidity in an interacting Fermi
gas~\cite{Hulet:science:2006:1,S.Jochim:Science302,Zwierlein:science:2006:1,Kinast:PhysRevLett.92.150402,Bourdel:PhysRevLett.93.050401}. One of the frontiers of this research~\cite{Jordens:Mott_transition,Schneider:2008Science322.1520S} is the realization of the Mott insulating limit for fermions in optical lattices~\cite{Lorenzo:PhysRevLett.101.210403,Scarola:PhysRevLett.102.135302}.

The most outstanding feature of this field is the amazing degree of control and tunability. While in solid state models like, e.g. the Hubbard model should be viewed only as approximate low-energy effective pictures with parameters hard to estimate and control, optical lattices allow to tune the interaction strength and other relevant parameters by simply changing the experimental setup. A well-known example is the possibility to control the value of the interatomic interactions by means of a magnetic field exploiting Fano-Feschbach resonances. This allows not only to study the evolution of the physics as a function of the ``standard'' control parameters (like the interaction or the particle density), but also to introduce new parameters which can strongly affect known phenomena or lead to entirely new quantum phases.
 
Binary mixtures of fermionic atoms with different masses (e.g. $^{6}$Li, $^{40}$K) introduce one of such additional parameters, namely the
difference between the hopping amplitudes of the two kinds of fermions.  If the two fermionic species hop with a different probability amplitude or, in more colloquial words, move at different velocities, familiar phenomena for equal-mass binary mixtures (which mimic spin-1/2 electrons in solids) can be modified and new physics may take place. The phase diagram at zero temperature has been worked out in the one-dimensional (1D) case~\cite{cazalilla:226402} for repulsive and attractive interactions and  for attractive interactions in the three dimensional case~\cite{dao:104517}. In both cases the focus has been on the broken-symmetry phases (superfluid, charge- and spin-ordered states), while this work is devoted to the so-called Mott transition, a transition from a fluid to an insulator driven by the local repulsion on a lattice populated by as many fermions as there are lattice sites (half-filled model).

The problem of two kinds of carriers with different hoppings is also reminiscent of solid-state systems in which two or more bands with different bandwidths are involved in the conduction. In this situation,
the possibility of an orbital-selective Mott transition, in which only one (or a few) band becomes insulating because of the electron-electron interaction, while another (or others) maintains a metallic behavior~\cite{ Anisimov_OSMT,Koga,Liebsch_OSMT_3,demedici_Slave-spins,Ferrero_OSMT,arita,ruegg,werner,vojta}
has been lively debated.

In this work, we will investigate the fate of such two-component imbalanced mixtures when the repulsive local interaction is switched on and gradually increased. Our knowledge about the limiting cases of a completely balanced mixture (the well-known Hubbard model) and the limit in which one species is completely localized (i.e. it has zero hopping), described by the so-called Falicov-Kimball model, already suggests that the evolution between these two situations will be far from trivial.

In the balanced case, the mechanism leading to the Mott transition has been elucidated by means of the dynamical mean-field theory (DMFT), the same method we use in this work. Within this approach, when the correlation strength is increased, the motion of the fermions gradually becomes more difficult, and the effective hopping is reduced. When the correlation reaches a critical value in which the hopping is renormalized to zero, the system becomes a Mott insulator. For interactions smaller than the critical value the system is always a Fermi-liquid, a normal fluid with well-defined long-lived excitations at low-energy, like standard metals in solid-state physics.
In the Falicov-Kimball model~\cite{Falicov:PhysRevLett.22.997} one of the fermionic species is not mobile. As a result, the mobile species too has a non-standard behavior and, even before the Mott transition occurs at large interactions, it does not behave like a normal Fermi-liquid. Therefore the evolution between the two limits should determine a destruction of the Fermi-liquid induced not only by the interaction, but also by the mass imbalance. 

The paper is organized as follows: In Sec.~\ref{sec:model}, we present the model and  we introduce the
dynamical-mean field theory method. In Sec.~\ref{sec:massrenorm}, we present the results for the quasiparticle renormalization and the phase diagram. In Sec.~\ref{sec:spectral}, we study the spectral functions for weak and strong couplings, while Sec.~\ref{sec:conclusion} is dedicated to concluding remarks.

\section{Model and Methods}
\label{sec:model}

We consider two-component fermionic mixtures loaded in a three-dimensional optical lattice, allowing the two species to have different masses, or, more generally, different hopping amplitudes. We will henceforth refer to the two species as ``light'' and ``heavy''.
Under conditions discussed, e.g. in Refs.~\onlinecite{Jaksch:2005AnPhy.315,werner:056401, duan:243202},
fermionic mixtures in optical lattices are described by a Hubbard model in which the lattice sites correspond to the minima of the optical potential:
\begin{equation}
  H=-\sum_{\langle i,j\rangle,\alpha}\,
  \ta(c^{+}_{i\alpha}c_{j\alpha}+\textrm{h.c.})\,+\,
  U\sum_{i}n_{i\ell}n_{i \textit{h}}.
\label{eq:hubbard}
\end{equation}
The index $\alpha$ refers to the light ($\ell$) and heavy ($\textit{h}$) fermionic
species ($\tl > \th$). The existence of interspecies Feschbach resonances between $^6$Li and $^{40}$K has been
demonstrated by Refs.~\onlinecite{Schreck:PhysRevLett.100.053201,Tiecke:PhysRevLett.104.053202}, and allows for
both an attractive or repulsive interaction with a tunable
strength, as assumed in (\ref{eq:hubbard}). 

In the following, we consider the case in which the number of fermions equals the number of lattice sites (half-filling) with a repulsive interaction. Under this condition, a Mott transition is possible as a function of $U$. 
Our focus will be to characterize how the Mott transition occurs for different values of the mass imbalance ratio which we define as $\zeta=(\tl-\th)/(\tl+\th)$, a quantity which goes  from $0$ (balanced system) to $1$ (Falicov-Kimball limit). We also define an average hopping $t = \frac{1}{2} (\tl+\th)$.

We focus on fluid states without considering the antiferromagnetic instability which is expected to occur at low-temperatures. Even if our solutions will not be representative of the actual groundstate of the model, they will properly describe the system at the finite temperatures at which the experiments can be carried out.
We notice in passing that the results for the attractive model can be obtained directly by
using a particle-hole transformation~\cite{dao:104517}. Here the Mott transition is mapped onto a pairing transition in which a Fermi liquid becomes an insulating state formed by localized ``pairs''~\cite{Keller_paring,Capone_paring,Toschi_paring}.

As mentioned above, we use DMFT, one of the most popular modern theoretical approaches
designed to treat  correlated fermions on a lattice. One of the main advantages of DMFT over other approaches is that it does not require any assumptions on the values of the coupling terms appearing in the Hamiltonian, and indeed becomes exact both in the small and in the strong interaction limit for any value of the imbalance $\zeta$. 

A practical implementation of DMFT requires the self-consistent
solution of a quantum impurity model, i.e. a model of a single-interacting
site coupled to a bath that allows for quantum
fluctuations on the correlated site. In the mean-field spirit the
site is representative of any site of the original lattice. This
correspondence is implemented via a self-consistency condition
which contains the information about the original lattice. More precisely the hybridization with the bath is parameterized by a frequency-dependent ``Weiss field'' $\hat{\mathcal{G}}_{\alpha}^{-1}(\iom) = \iom + \mu -\Delta_{\alpha}(\iom)$, where $\Delta_{\alpha}(\iom)$ is the hybridization function.
The general form of the self-consistency equation (we write it for
simplicity for the normal fluid phase, but its generalization
to the broken-symmetry phases is straightforward) is
\begin{equation}
\label{eq:sc}
{\mathcal{G}}_{\alpha}(\iom) = \int \mathrm{d}\varepsilon
\frac{N_{\alpha}(\varepsilon)}{\iom + \mu_{\alpha} -
\varepsilon-\Sigma_{\alpha}(\iom)}+\Sigma_{\alpha}(\iom),
\end{equation}
where $\Sigma_{\alpha}(\iom) =
{\cal{G}}^{-1}_{\alpha}(\iom)-G^{-1}_{\alpha}(\iom)$ is the local
self-energy, $N_{\alpha}(\varepsilon)$ is the non-interacting
density of states. 

In this work we use a semicircular density of states
$N_{\alpha}(\varepsilon) = \frac{1}{\sqrt{2\pi\Da}}\sqrt{\Da^2-\varepsilon^2}$ with half-bandwidth
$\Da$, for which (\ref{eq:sc}) is greatly simplified and
becomes
\begin{equation}
\label{sc_bethe} {\mathcal{G}_{\alpha}}(\iom) = \iom +\mu_{\alpha}
-\frac{D_{\alpha}^2}{4} G_{\alpha}(\iom).
\end{equation}

The above density of states has been shown to satisfactorily reproduce results in three spatial dimensions with the half-bandwidth related to the nearest-neighbor hopping in three dimensions by the relation $\Da = 6\ta$. We can obviously define an average half-bandwidth $D = \frac{1}{2} (\Dl+\Dh)$, and observe that the definition of $\zeta$ can be reexpressed in terms of $\Da$ as  $\zeta=(\Dl-\Dh)/(\Dl+\Dh)$. For more details on DMFT, we refer to Ref.~\onlinecite{georges:rmp:1996:1}.

\section{Mass renormalization}
\label{sec:massrenorm}

In very general terms, the main effect of the interparticle interaction $U$ is to reduce the mobility of the fermions.
This phenomenon, which eventually leads to the Mott localization, is measured by the effective mass of the carriers $m^*$. Within DMFT $m^*$  is the inverse of the quasiparticle weight defined by $Z^{-1}_{\alpha}=\Big(1-\mathrm{d}\,\textrm{Re}[\Sigma_\alpha(\omega)] / \mathrm{d} \omega\Big|_{\omega=0} \Big)$. For a non-interacting system $\Za =1$, while the localization is associated to a vanishing $\Za$ corresponding to an infinite effective mass. A small value of $\Za$ is the signature of a highly-correlated quantum fluid. Here, because the two species have different bare masses and hoppings, a different renormalization is expected and the outcome is not obvious. 

In the case of two different bands in a solid, both carrying up and down electrons, it is natural that the heavy band, which has a smaller $t$ is more affected by the interaction than the light one. Therefore the former will be more renormalized than the latter, and the ratio between the renormalized hoppings $(\Zl\tl )/(\Zh\th )$ will be larger than that of the bare hoppings $\tl/\th$. This opens the way to a possible orbital-selective Mott transition in which the renormalized heavy-band hopping goes to zero, while the light one remains metallic. The actual realization of the transition depends on several factors, but the qualitative behavior of the renormalization factors universally follows the above expectations~\cite{ Anisimov_OSMT,Liebsch_OSMT_3,demedici_Slave-spins,Ferrero_OSMT}.

In principle this effect (heavy fermions being more sensitive to the correlations) can take place also in our two-component mixture, but another effect competes with it. When the system is at half-filling (or close) and the correlation is strong, we essentially have one fermion on each lattice site, even in the itinerant state just before a Mott transition. In this situation, in order to move one light fermion, we are bound to move it on a site which is already populated by a heavy one (Pauli principles forbids to have two fermions of the same species on the same site) and then move the heavy one on the site previously occupied by the moved light fermion. This means that, in order to move the light fermions, we are forced to move the heavy ones as well. When we are in this regime, we expect the renormalization factors $\Za$ to compensate the hopping imbalance. In other words we expect $\Zl < \Zh$, in contrast with the case of two bands in a solid. Even more simply, close to the Mott transition, the motion of each carrier is correlated to that of any other, and there is no energy gain in having one species moving faster than the other. Our DMFT analysis will prove that this latter effect is the dominant one.

\subsection{Linearized DMFT}

\begin{figure}[!ht]
  \begin{center}
    \includegraphics[width=8cm,height=6cm,clip=true]{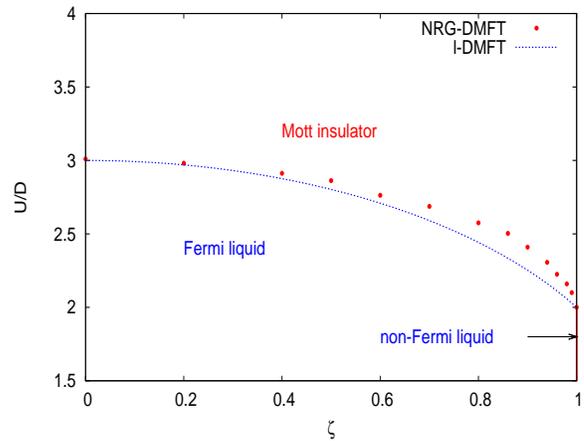}
  \end{center}
  \caption{(Color online) Phase diagram of the Mott insulator - fluid transition.}
\label{fig:Mott_transition}
\end{figure}

Before presenting the full numerical solution of DMFT and discussing the quasiparticle properties in details, we show analytical results for the critical interaction $U_c$ obtained using a further approximation which has proved reliable in the balanced Hubbard model,  the linearized DMFT approximation (l-DMFT)~\cite{Potthoff:2000EPJB.13.257B}. This method approximates the DMFT Weiss field  with a single-pole function and it is expected to be accurate close to the Mott transition. For details on this approximation, we refer the reader to the original paper~\cite{Potthoff:2000EPJB.13.257B}. Let us summarize here the main results for the mass-imbalanced case. Within l-DMFT the hybridization function is simply given by
\begin{equation}
\Delta_\alpha(\omega)=\frac{\Va^2}{\omega}, 
\end{equation}
where  $\Va$ is the hybridization between the approximate bath and the impurity. 
Close to the critical $U = U_c$ for the Mott transition we can approximate the quasiparticle weights to second order in $\Va/U_c$:
\begin{eqnarray}
\Zl= (16 \Vl \Vh + 16 \Vh^2 + 4 \Vl^2)/U_c^2,\label{eq:Zl}\\ 
\Zh= (16 \Vh \Vl + 16 \Vl^2 + 4 \Vh^2)/U_c^2\label{eq:Zh}. 
\end{eqnarray}
For a semicircular DOS the DMFT self-consistency condition for each species implies~\cite{Potthoff:2000EPJB.13.257B}
\begin{eqnarray}
\Va^2=\frac{\Da^2}{4} \Za,\label{eq:sccl}
\end{eqnarray}
Eqs.~(\ref{eq:Zl}),(\ref{eq:Zh}) and~(\ref{eq:sccl}) lead to a quartic equation for $\Vl/\Vh$:
\begin{equation}
\label{eq:cu}
\left(\frac{\Dl}{\Dh}\right)^2\left(\frac{\Vl}{\Vh}+2\right)^2=4\left(\frac{\Vl}{\Vh}\right)^2\left(\frac{\Vl}{\Vh}+\frac{1}{2}\right)^2,
\end{equation}
which has a single physically-relevant solution
\begin{equation}
\label{eq:solc}
\Vl=\frac{\Vh}{4}\left(\frac{\Dl}{\Dh}-1+\sqrt{1+14\frac{\Dl}{\Dh}+\left(\frac{\Dl}{\Dh}\right)^2}\right),
\end{equation}
implying, using Eq.~(\ref{eq:sccl}),
\begin{equation}
\label{eq:zrat}
\frac{\Zl}{\Zh} = \frac{\Dh^2}{\Dl^2}\left[\frac{1}{4}\left(\frac{\Dl}{\Dh}-1+\sqrt{1+14\frac{\Dl}{\Dh}+\left(\frac{\Dl}{\Dh}\right)^2}\right)\right]^2.
\end{equation}
Replacing $\Vl$ from Eq.~(\ref{eq:solc}) in Eq.~(\ref{eq:sccl}) and using that $\Dl=D(1+\zeta)$, $\Dh=D(1-\zeta)$, where $D = 1/2 (\Dl + \Dh)$ is the average bandwidth, we obtain the following expression for the critical interaction:
\begin{equation}
\label{Uc_lDMFT}
U_c(\zeta)= D \sqrt{-3 \zeta^2+2 \sqrt{4-3 \zeta^2}+5}.
\end{equation}

$U_c$ is a monotonically decreasing function of $\zeta$ which continuously connects the result for species of identical mass
$U_c(0) = 3D$ to $U_c(1)= 2D$, which is the result for the Falicov-Kimball limit in which one of the species is localized. At fixed average hopping $t$, the larger the difference in the bare mass, the easier it is to localize the system. The l-DMFT result for the critical interaction as a function
of $\zeta$ is shown in Fig.~\ref{fig:Mott_transition} as a dashed line.

\subsection{Full DMFT calculations}

We now turn to the full DMFT solution. We use two different impurity solvers, the numerical renormalization group (NRG)~\cite{Bulla_nrg}, which is particularly accurate for low-energy features, and the exact diagonalization (ED)~\cite{edpapers}, an unbiased method which however introduces some approximation to the spectra being represented as a sum of discrete poles (here we approximated the bath with seven discrete levels). Both methods work at zero temperature. 
In the NRG calculations we set the discretization parameter $\Lambda=2$ and we keep 500 states at each NRG iteration. The spectral densities are computed following Refs. \onlinecite{bulla1,bulla2}. In the ED solution we discretize the bath using a total number of levels $N_s=9$ having verified that this number is sufficient to obtain converged results.

\begin{figure}[!ht]
  \begin{center}
    \includegraphics[width=8cm,clip=true]{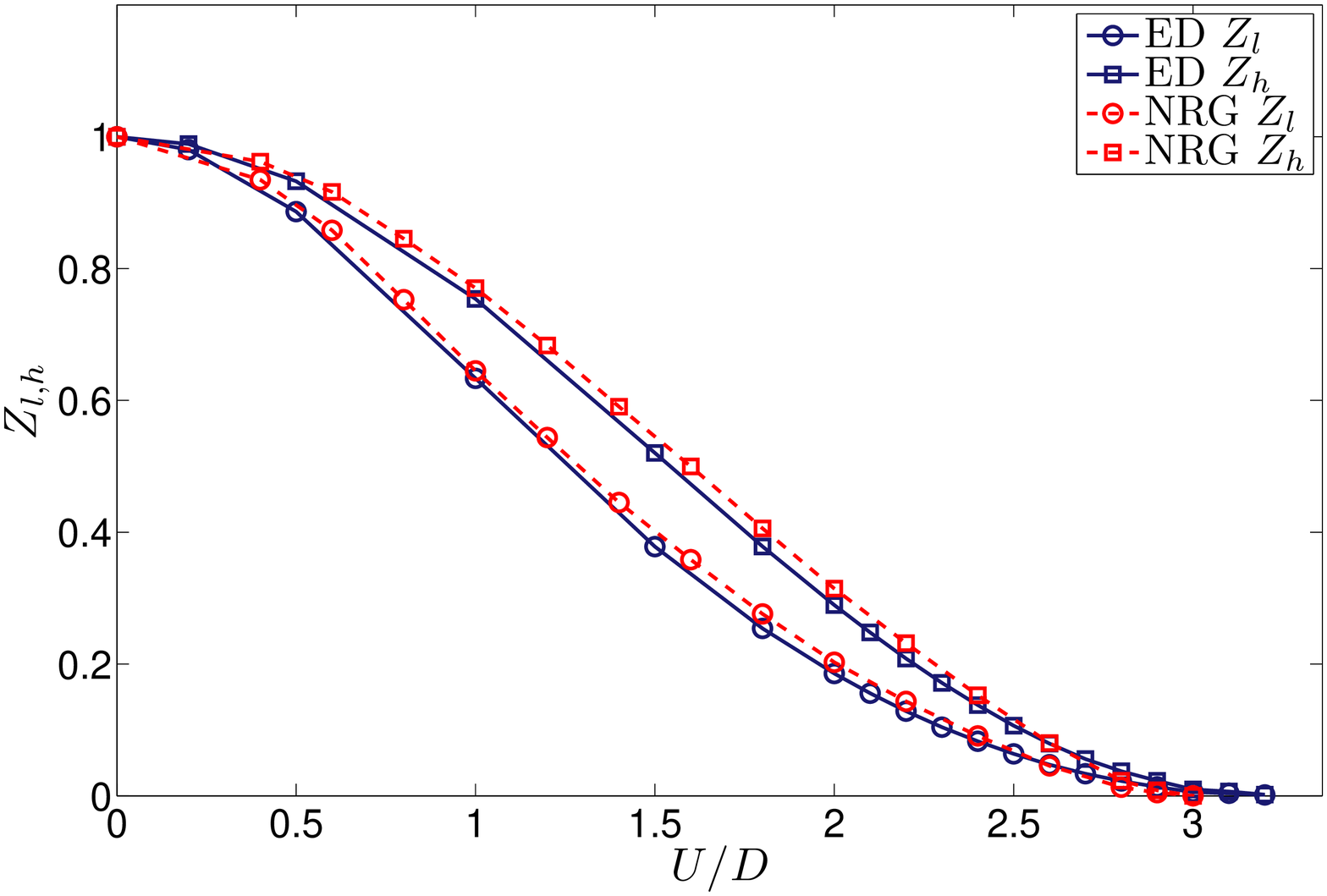}\\
    \includegraphics[width=8cm,clip=true]{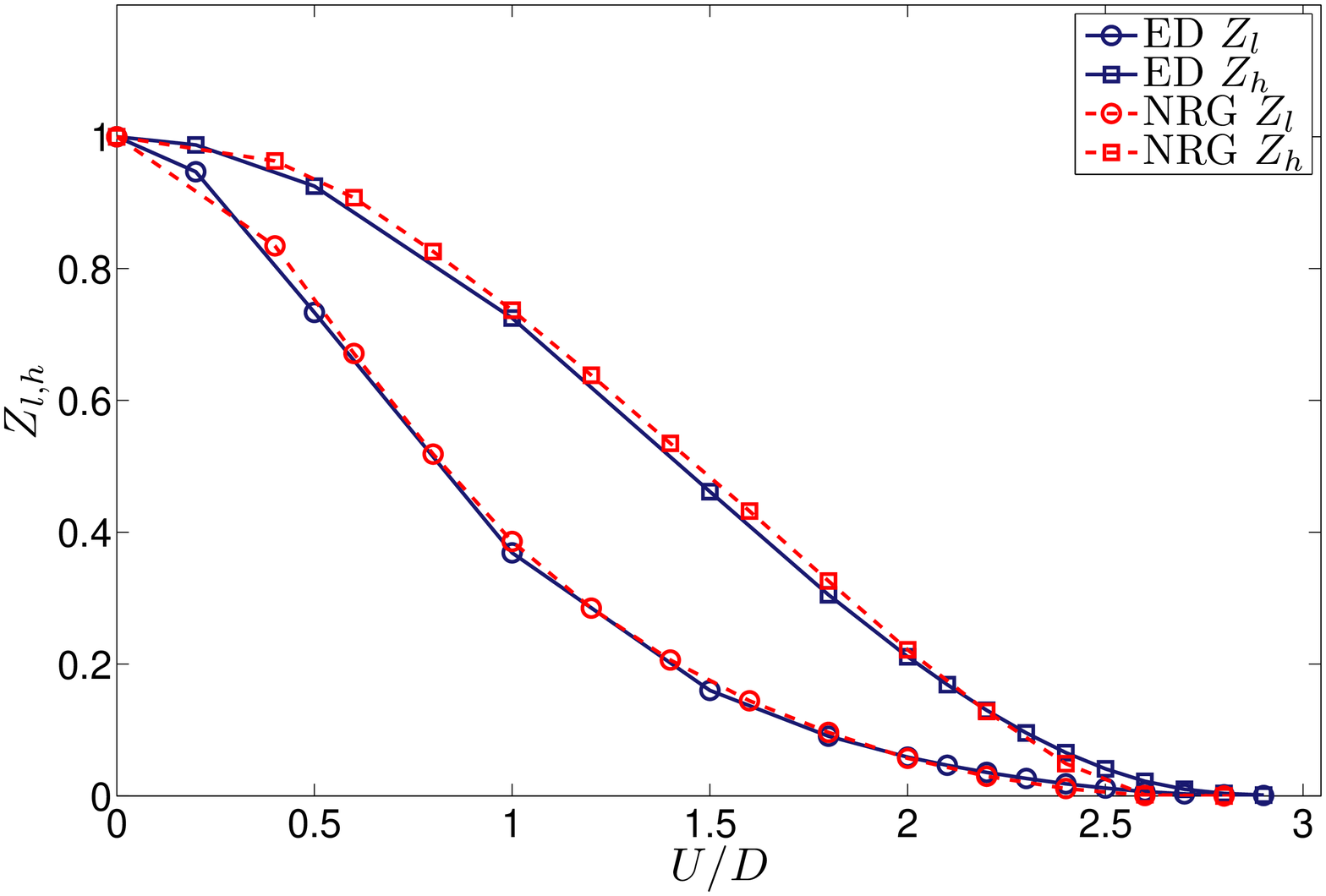}
  \end{center}
  \caption{(Color online) Quasi-particle weight $\Za$ at the Fermi level as a function of the
           interaction strength $U/D$ for $\zeta=(\tl-\th)/(\tl+\th)=0.4$ (upper panel)
           and $0.8$ (lower panel).}
\label{fig:EDvsNRG_z}
\end{figure}

Fig.~\ref{fig:EDvsNRG_z} shows the quasiparticle weight of both
species for two different mass imbalances $\zeta=0.4, 0.8$. The very good agreement between ED and NRG represents a non-trivial test for the accuracy of our calculations. Both for the small imbalance and the strong imbalance cases we find that the light fermions are more renormalized (i.e. they have a smaller $Z$) compared to their heavy partners, even if the effect is much more pronounced for $\zeta = 0.8$. Our solution therefore shows that the second physical effect that we described above prevails already for moderately imbalanced mixtures: When two species with different mobility are mixed, the interactions tend to balance their properties. It is however important to underline that the reduction of $\Zl$ does not perfectly balance the renormalized hoppings and the effect is far from being trivial (as already suggested by Eq.~\ref{eq:zrat}).

Repeating the calculations for different values of $\zeta$ we constructed the phase diagram for the Mott transition from a fluid to an insulator in the $U,\zeta$ plane, which we report in Fig.~\ref{fig:Mott_transition} together with the analytical formula (\ref{Uc_lDMFT}) from l-DMFT. The numerical result obtained with full DMFT (ED and NRG give indistinguishable results) are slightly above the l-DMFT predictions, but follow a very similar behavior, which bridges between the value for the balanced model $U \simeq 3D$ and that of the Falicov-Kimball model $U = 2D$. As we will discuss in the following, the metallic region below $U_c$ is always a Fermi liquid except for the Falicov-Kimball limit $\zeta =1$, where the Fermi-liquid picture breaks down. Nonetheless, moving along the $\zeta$ axis, there is a continuous reduction of the coherence of the metallic phase.

\section{Spectral functions}
\label{sec:spectral}

In the previous section we have analyzed how the effective mass of the two species of fermions evolves as a function of $U$ and we have drawn a phase diagram in the imbalance-correlation plane which highlights how the fluid state turns into a Mott insulator. We now extend our analysis to the spectral functions, which contain information about the excitation spectrum for each species. This is a key quantity to characterize the nature of the itinerant states and the approach to the insulating state. In particular, we can identify if the system presents long-lived  low-energy excitations and the characteristic energy scale below which the fermionic motion is coherent despite the interaction between the species. Most interestingly, spectral functions are experimentally accessible  in trapped cold atomic systems~\cite{Li_K_mixture} via radio-spectroscopy~\cite{Grimm2004_RF} or Raman spectroscopy~\cite{DaoCarusotto2007,StewartJin2008}.

We have therefore computed $A_{\alpha}(\omega) = -1/\pi \textrm{Im} G_{\alpha}(\omega + i0^+)$ within DMFT using NRG as impurity solver\cite{bulla1,bulla2}. Indeed, this method allows for an arbitrarily-fine resolution at low frequency, where the most interesting physics will take place.
We will show that for all values of $\zeta$ the system is a normal Fermi liquid for $U  < U_c(\zeta)$, even if the coherence energy scale can be very small and, more interestingly, highly species-dependent. Due to the momentum-independence of the self-energy within DMFT, a Fermi-liquid behavior implies that the spectral function at $\omega=0$ is pinned to its non-interacting value~\cite{georges:rmp:1996:1}. For our model at half-filling, this implies $A_{\alpha}(0) = 2/(\pi \Da)$ for both species. For this reason, we always plot $1/2\pi \Da A_{\alpha}(\omega)$, a quantity which goes to one for $\omega=0$ when the system is a Fermi liquid.

\subsection{Weak-coupling regime}

We start our analysis from the weak-coupling regime, choosing $U=0.5D$ for the sake of definiteness. For small values of the imbalance, the system is expected to behave like a standard Fermi liquid with slightly different renormalizations for the two species. On the other hand, the large-imbalance limit can provide non-standard physics even when the correlation strength is moderate. Therefore we only focus on relatively large values of $\zeta$ from 0.8 to 0.9998.

\begin{figure}[!ht]
  \begin{center}
  \includegraphics[width=8cm,height=6cm,clip=true]{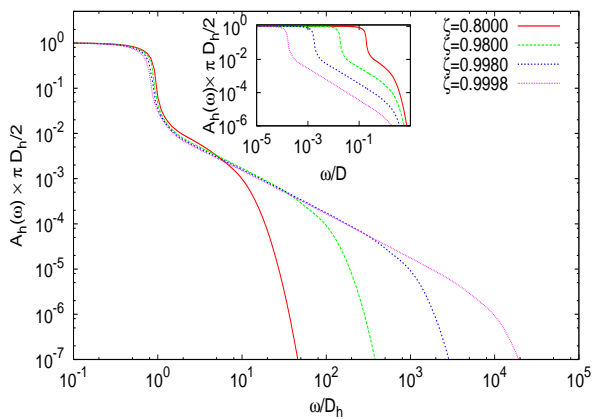}
  \end{center}
  \caption{(Color online) Spectral functions of the heavy fermions for $U=0.5D$ and
           $\zeta=0.8, \, 0.98, \, 0.998, \, 0.9998$. The $\omega$-axis is renormalized by $\Dh$
           and the spectral functions $A_{\textit{h}}$ are multiplied by $1/2\pi \Dh$.
           The inset shows the same quantities as a function of $\omega/D$.}
\label{fig:AhU0.5}
\end{figure}

In the Falicov-Kimball limit corresponding to $\zeta = 1$, the heavy fermions are frozen in the lattice whereas the light fermions can move. This case can be mapped onto a well-known solid-state problem, the absorption of  x-rays in a metal, for which we refer to Refs.~\onlinecite{Pablo:PhysRevB.75.115112,Anders:PhysRevB.71.125101}. Within this mapping, the properties of the heavy fermionic species correspond to the core level of the metal, whose spectral function has a distinctive power-law behavior at low frequency~\cite{Si:PhysRevB.46.1261}:
\begin{equation}
A_{\textit{h}}(\omega)\propto |\omega|^{(\alpha-1)}.
\end{equation} 
Our DMFT results show that this result still holds over a given frequency range at finite values of the imbalance and weak values of the interaction. The accuracy of NRG-DMFT calculations in the evaluation of the power-law behavior has been demonstrated, e.g., in Ref. \cite{Pablo:PhysRevB.75.115112} through a comparison with exact results.

Fig.~\ref{fig:AhU0.5} shows the spectral functions of the heavy species for
different hopping imbalances $\zeta=0.8,0.98,0.998,0.9998$.  In this figure, we
have rescaled both axes in order to have a suitable comparison between spectral
functions corresponding to different $\zeta$. The energy $\omega$ is
renormalized by $\Dh$ while the spectral density
$A_{\textit{h}}(\omega)$ is multiplied by a factor $1/2\pi \Dh$.  In our rescaled
units, all these spectral functions have a similar behavior.  Starting from high
energy, they first enter a region where they all have \emph{the same} power-law
behavior at an energy which is larger the larger $\zeta$. Fitting our
data for all $\zeta$ gives the same numerical value for the power-law exponent
$\alpha-1=-0.94$ and we have that $A_{\textit{h}}=0.0137(\omega/D)^{-0.94}$.
This value is remarkably consistent with the exponent found in the
Falicov-Kimball model $\alpha_{FK}-1=-0.9375$.  At a lower energy scale
$\omega_p \sim \Dh$ the spectra eventually deviate from this power-law behavior
and construct a low-energy peak. The reason for this peak is that the heavy
fermions are not completely localized.  At an energy scale smaller than
$\omega_p$ (which corresponds to their kinetic energy) the heavy fermions behave
like well-defined mobile quasiparticles. Indeed, at $\omega=0$, the spectral
functions go to the same finite value $A_{\textit{h}}(\omega=0)=2/(\pi \Dh)$
which is the typical result for a Fermi-liquid within DMFT.

\begin{figure}[!ht]
  \begin{center}
    \includegraphics[width=8cm,height=6cm,clip=true]{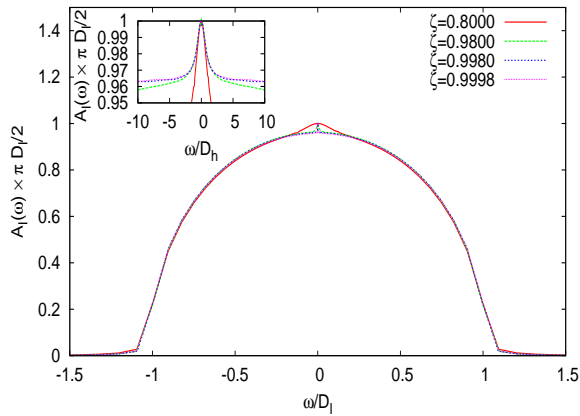}
  \end{center}
  \caption{(Color online) Spectral functions of the light species for different values of the
           mass imbalance $\zeta=0.8, \, 0.98, \, 0.998, \, 0.9998$. The energy $\omega$ is
           renormalized by $\Dl$ and the spectral density
           $A_{\textit{l}}(\omega)$ is multiplied by a factor $1/2 \pi \Dl$. The inset
           shows the very-low energy part rescaling the energy axis by $\Dh$ instead of
           $\Dl$ in order to show the effect of the heavy species on the light one.}
\label{fig:AlU0.5}
\end{figure}

We now turn to the light species. Fig.~\ref{fig:AlU0.5} shows the spectral
functions of the light species for the considered values of $\zeta$. In order
to obtain a suitable comparison, we rescale again the energy by 
$\Dl$. We remark that, interestingly, the spectral functions of the light
species $A_{\textit{l}}(\omega)$ also have different behaviors corresponding to
the same energy scales we discussed for the heavy species. In the energy regime
$\omega > \omega_p \sim \Dh$, all spectral functions collapse onto the same curve.
At these energies, the spectral
function of the light species behaves as if the heavy particles had an infinite
mass and follows the behavior expected for a Falicov-Kimball system. Hence,
all $A_{\textit{l}}(\omega)$ have the same behavior when we renormalize the
energy scale by the hopping parameter. By contrast, at low energy
($\omega < \omega_p$), the finite mass of the heavy particles becomes apparent
and the spectral function develops a peak structure.
If we rescale the energy $\omega$ by the hopping of the {\it heavy} fermions
$\Dh$, the low-energy part of the different spectra again collapse on a single
curve which can be nicely fitted by a Lorentzian (see the inset of
Fig.~\ref{fig:AlU0.5}). This implies that the width of these Lorentzian peaks
is proportional to $(1-\zeta) \propto \Dh$.  This result can be understood as
follows: At low energy, the nature of the correlations between heavy and light
species changes. While at high energies the interaction between species
destroys the coherence of the light species and generate a cascade of
particle-hole excitations in the spectrum of the heavy species, at low energies
they generate a new Fermi liquid with a renormalized mass for both light and
heavy species. This shows that the typical energy scale $\omega_p \propto
(1-\zeta)$ introduced above is also the scale at which the nature of the
correlations between the two species changes.

\subsection{Strongly-correlated regime}

We now consider the strong coupling regime with interaction strength $U=2D$. We choose this value because it is below the Mott transition point for any value of $\zeta < 1$, but is also coincides with the limiting value of $U_c$ in the Falicov-Kimball limit $\zeta =1$. Therefore we will always be in a liquid state, even if increasing $\zeta$ will drive the system close to a Mott transition (see Fig.~\ref{fig:Mott_transition}). In order to investigate how the effect of correlations increases when $\zeta$ gets bigger, we span a wide range of $\zeta$ from small to very large imbalance.

\begin{figure}[!ht]
  \begin{center}
    \includegraphics[width=8cm,height=6cm,clip=true]{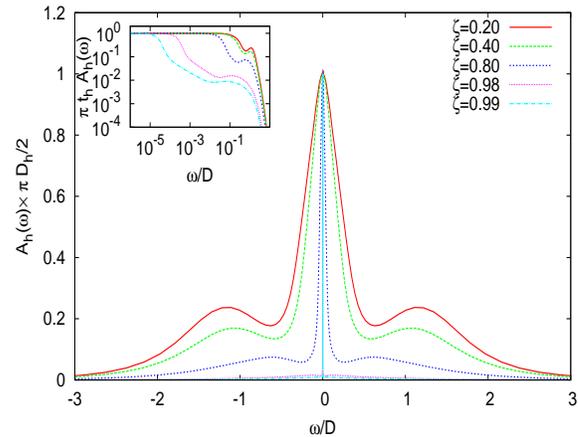}
  \end{center}
  \caption{(Color online) Spectral functions of the light species for $U=2D$ and
           $\zeta=0.2, \, 0.4, \, 0.8, \, 0.98, \, 0.99$. All spectral functions are multiplied by
           $1/2\pi \Dl$.}
\label{fig:Adw_U2}
\end{figure}

\begin{figure}[!ht]
  \begin{center}
    \includegraphics[width=8cm,height=6cm,clip=true]{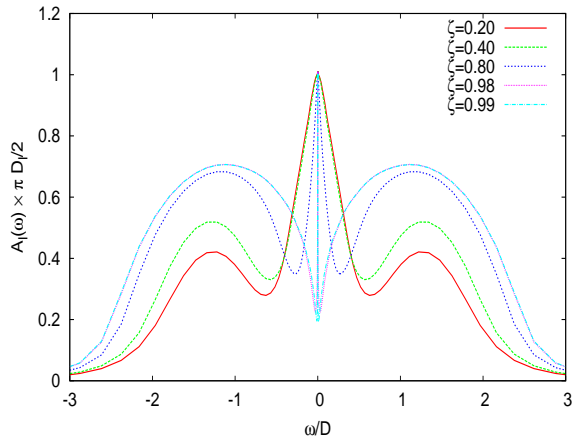}
  \end{center}
  \caption{(Color online) Spectral functions of the heavy species for $U=2D$ and
           $\zeta=0.2, \, 0.4, \, 0.8, \, 0.98, \, 0.99$. All spectral functions are multiplied by
           $1/2\pi \Dh$. In the inset, the same quantities are plotted on a log-log scale.}
\label{fig:Aup_U2}
\end{figure}

Fig.~\ref{fig:Adw_U2} and Fig.~\ref{fig:Aup_U2} present the spectral densities
of the heavy fermions and light fermions for different values of the mass imbalance
$\zeta=0.2, 0.4, 0.8. 0.98, 0.99$. In order to check the Fermi-liquid
property, we multiply $A_{\alpha}(\omega)$ by $1/2\pi \Da$. Again, the spectral
functions behave as a normal Fermi liquid with $1/2\pi\Da A_{\alpha}(0)=1$ for
every value of $\zeta$. Unlike the weak-interaction limit, for large values of
$U$ the behavior of these spectral functions strongly depends on the hopping
imbalance $\zeta$. At high energy $\omega \sim U$, the physics of both species
is dominated by the interaction that induces the formation of Hubbard bands with
incoherent excitations around $\omega \pm U/2$.

The evolution of the light species as a function of $\zeta$ shows the typical approach to a Mott transition, even if the driving parameter is not the interaction.  We are, in fact, moving along a horizontal line pointing toward the Mott transition at $\zeta=1$ in the phase diagram of Fig.~\ref{fig:Mott_transition}. For small $\zeta=0.2$ a sizable peak around $\omega = 0$ lives between the Hubbard bands at $\omega = \pm U/2$. When increasing $\zeta$, this peak shrinks (its width is proportional to $\Zl \tl$) and the spectral weight is transferred to the Hubbard bands. Very close to $\zeta=1$ the peak becomes narrow and eventually vanishes in the limit $\zeta=1$. The main qualitative difference with respect to the Mott transition obtained in the balanced model as a function of the interaction strength is that in the present case there is no separation between the metallic peak and the Hubbard bands, and the gap opens continuously at the metal-insulator transition. We remind that in the balanced model a precursor of the Mott gap opens already in the metallic phase, so that the metallic peak is separated from the Hubbard bands and the gap at the transition point is already finite.
The continuous opening of the gap when $\zeta$ increases is however not surprising because in the limit $\zeta=1$ one has to recover the Falicov-Kimball model which shows a continuous opening of the gap at the metal insulator transition.

An even richer evolution is shown by the heavy species, in which Mott and
Falicov-Kimball physics are entangled. At small hopping imbalance
($\zeta=0.2$), $A_{\textit{h}}(\omega)$ has the typical shape of a
highly-correlated Fermi liquid with a peak at zero energy flanked by the
Hubbard bands.  Close to the $\zeta=1$ limit, the behavior of
$A_{\textit{h}}(\omega)$ resembles that of the non-Fermi liquid found in the
Falicov-Kimball model, except at very low energies. There, the Fermi-liquid
behavior is recovered in the same way as we discussed in the weak-coupling
regime (compare the insets of Fig.~\ref{fig:Aup_U2} and Fig.~\ref{fig:AhU0.5}).
For large $\zeta$, we obtain again a power-law behavior in the
intermediate-energy regime. However, at variance with the weak-coupling regime,
the exponent is strongly dependent on $\zeta$. As shown in the inset of
Fig.~\ref{fig:Aup_U2}, $\alpha$ decreases when $\zeta$ tends to $1$. Moreover,
the characteristic energy $\omega_{p*}$ which determines the return to
Fermi-liquid behavior is no longer a linear function of $(1-\zeta)$. We obtain
$\omega_{p*}$ about $10^{-2}D$ for $(1-\zeta)=2\times 10^{-1}$, $10^{-4}D$ for
$(1-\zeta)=2\times 10^{-2}$) and $10^{-8}D$ for $(1-\zeta)=2\times 10^{-3}$.
The scale below which the system behaves as a coherent Fermi liquid is
therefore much smaller than in the weak-interaction limit. Such a reduction of
the coherence scale for the heavy species can be understood in terms of the
Mott physics. When $\zeta$ approaches $1$, the system is closer to the Mott
transition, hence the quasiparticle weight $\Zh$ is strongly reduced. This
implies that the width of the quasiparticle part $\omega_{p*}$ must be
proportional to $\Zh \Dh$, much smaller than $\Dh$. Only at weak coupling, when
$\Zh$ is close to $1$ and weakly $\zeta$-dependent (i.e. the mass
renormalization effect is negligible), do we obtain $\omega_{p} \sim \Dh$.

We have shown that the interactions (Mott physics) have an influence on the
power-law behavior of the spectra. Conversely, at high energy, the power-law
behavior of the spectral function has consequences for the details of the Mott
transition. Indeed, the Hubbard bands in the spectral functions of the heavy
species are not exactly located at $\pm U/2$, as in the case of light fermions
or in the balanced Hubbard model. For small values of $\zeta=0.2,\, 0.4$, the
power-law behavior is less pronounced and the Hubbard bands are close to $\pm
U/2$. But for a high hopping asymmetry $\zeta>0.8$, the power-law behavior
becomes dominant and transfers spectral weight to low energies. Consequently,
the Hubbard bands are pushed closer to the Fermi level, as shown in
Fig.~\ref{fig:Adw_U2}.

\section{Conclusion}
\label{sec:conclusion}

We have studied the Mott transition of mixtures of fermionic species with
different masses in a cubic optical lattice with repulsive on-site interaction.
We have shown the dependence of the critical value of the interaction $U_c$ for the
Mott transition on the hopping imbalance $\zeta$ both via analytical and numerical
techniques.  Our first result concerns the mass renormalization observed for
both components in the mixture when they enter the strong-interaction regime.
Interestingly, the light species is more renormalized than the heavy one, so
that the effective masses induced by the interaction are closer than the bare
masses. The two species tend to move at the same velocity before a Mott
transition localizes them simultaneously. Note that an experimental study on
the effective masses of a Fermi gas with {\it density} imbalance has been reported in
Ref.~\onlinecite{NatureCSalomon}.

Our second main result is a thorough characterization of the spectral properties of 
light and heavy fermions as a function of the interaction strength and of the hopping imbalance. Here 
we observe a  variety of deviation from the standard behavior of correlated Fermi gases 
which range from heavy fermion behavior to a power-law dependence on the frequency.

The experimental test of our predictions can be performed by conventional and successful
techniques in the ultracold atom field such as the radio-spectroscopy~\cite{Grimm2004_RF} or the Raman
spectroscopy~\cite{DaoCarusotto2007,StewartJin2008}.

However, the experimental detection of the highly incoherent state we predict is
not simple, at least with present experimental set-up. 
Let us consider for example the behavior of the heavy species for sizeable $U$ shown in Fig. 6. Here the system
is in principle a Fermi liquid at zero temperature for every value
of $\zeta$, but, especially at large values of $\zeta$, the Fermi-liquid behavior is associated with a very narrow
quasiparticle peak which is expected to be rapidly washed out at finite temperature exceeding the small
coherence temperature scale. Therefore the experimental detection of this peak is very hard in  the present experimental 
situation, where the temperature is hardly lower than a sizeable fraction of the Fermi energy $T/T_F \simeq 0.15$.

However, the fingerprints of our anomalous
state can be observed even when the temperature washes out the narrow coherent peak 
since the spectral function will be completely different from a standard Fermi liquid. In particular
the spectral function will show a dip at zero energy corresponding to an incipient localization of the 
carriers even if the system is not fully gapped, a feature which can be reasonably observed in present
experiments.
A further obstacle to the detection of the incoherent metallic
state comes from the competition with the antiferromagnetic state. At low temperature the
system is expected to antiferromagnetically order at least in bipartite lattice without 
frustration.  Nevertheless, the physics associated to the Mott localization of the fermions can still be observed 
at finite temperatures above the N\'eel temperature, as it has been demonstrated in the field of solid state, 
where, for example, the finite temperature properties of V$_2$O$_3$ have been shown to follow the behavior
predicted by DMFT once the temperature exceeds the N\'eel temperature\cite{limelette}.

We finally note that an ideal candidate for an experimental check, the
$^6$Li/$^{40}$K mixture, has been realized with well controllable interspecies
Feschbach resonance~\cite{Li_K_mixture}. A simple numerical estimate shows that
the hopping imbalance $\zeta$ can be varied over a large range by changing the
lattice depth $V_0/E_R$ ($\zeta\ll 1$ at small $V_0/E_R$ and $\zeta\simeq 0.9$
for $V_0/E_R\simeq 15$), realizing the regimes of $\zeta$ where we
identified the most striking anomalies both in the quasiparticle
renormalization and in the spectral properties.

\begin{acknowledgments}
M. C. acknowledges financial support of the European Research Council through
FP7/ERC Starting Independent Research Grant ``SUPERBAD'' (Grant Agreement
240524). T.-L. D. acknowledges the computation support from the group Optic
Atomic of LCFIO for numerical simulations and the financial support of IFRAP
and RTRA. P. S. C. acknowledges financial support from PIP 11220080101821 of
CONICET.
\end{acknowledgments}



\end{document}